\documentclass[
 amsmath,amssymb,
 aps,
pra,
reprint
]{revtex4-2}

\usepackage{graphicx}
\usepackage{dcolumn}
\usepackage{bm}
\usepackage{physics}
\usepackage{amsfonts, amsmath}
\usepackage{dsfont}
\usepackage{amsthm}
\usepackage{bm}
\usepackage{hyperref}
\usepackage{bbold}
\usepackage{color}
\usepackage{lipsum,babel}
\usepackage[normalem]{ulem}
\usepackage{tabularx}

\newcommand{\poly}{\text{poly}}

\newtheorem{theorem}{Theorem}
\newtheorem{corollary}{Corollary}

\newtheorem{lemma}{Lemma}

\begin{document}
\definecolor{navy}{RGB}{46,72,102}
\definecolor{pink}{RGB}{219,48,122}
\definecolor{grey}{RGB}{184,184,184}
\definecolor{yellow}{RGB}{255,192,0}
\definecolor{grey1}{RGB}{217,217,217}
\definecolor{grey2}{RGB}{166,166,166}
\definecolor{grey3}{RGB}{89,89,89}
\definecolor{red}{RGB}{255,0,0}

\preprint{APS/123-QED}

\title{Classical simulability of constant-depth linear-optical circuits with noise}
\author{Changhun Oh}
\email{changhun0218@gmail.com}
\affiliation{Department of Physics, Korea Advanced Institute of Science and Technology, Daejeon 34141, Korea}

\begin{abstract}
Noise is one of the main obstacles to realizing quantum devices that achieve a quantum computational advantage.
A possible approach to minimize the noise effect is to employ shallow-depth quantum circuits since noise typically accumulates as circuit depth grows.
In this work, we investigate the complexity of shallow-depth linear-optical circuits under the effects of photon loss and partial distinguishability.
By establishing a correspondence between a linear-optical circuit and a bipartite graph, we show that the effects of photon loss and partial distinguishability are equivalent to removing the corresponding vertices.
Using this correspondence and percolation theory, we prove that for constant-depth linear-optical circuits with single photons, there is a threshold of loss (noise) rate above which the linear-optical systems can be decomposed into smaller systems with high probability, which enables us to simulate the systems efficiently.
Consequently, our result implies that even in shallow-depth circuits where noise is not accumulated enough, its effect may be sufficiently significant to make them efficiently simulable using classical algorithms due to its entanglement structure constituted by shallow-depth circuits.
\end{abstract}

\maketitle

\section{Introduction}
Quantum optical platforms using photons are expected to play versatile roles in quantum information processing, such as quantum communication, sensing, and computing~\cite{duan2001long, kok2007linear, sangouard2011quantum, pirandola2018advances, slussarenko2019photonic, bourassa2021blueprint}.
Especially, quantum optical circuits using linear optics are more experimentally feasible and still have the potential to provide a quantum advantage for quantum computing; representative examples are boson sampling~\cite{aaronson2011computational} or Knill-Laflamme-Milburn~(KLM)~protocol for universal quantum computation~\cite{knill2001scheme}.

However, as in other experimental platforms, one of the main obstacles to implementing a large-scale quantum device to perform interesting quantum information processing is noise; especially, photon loss and partial distinguishability of photons in photonic devices are typically the most crucial noise sources~\cite{zhong2020quantum, zhong2021phase, madsen2022quantum, deng2023gaussian}.
Through experimental realizations of intermediate-scale quantum devices using photons and thorough theoretical analysis of the effect of loss and noise, many recent results show that they can significantly reduce the computational power of the quantum devices~\cite{zhong2020quantum, zhong2021phase, madsen2022quantum, deng2023gaussian, oszmaniec2018classical, renema2018classical, renema2018efficient, garcia2019simulating, qi2020regimes, renema2020simulability, oh2021classical, shi2022effect, oh2023classical, liu2023complexity, oh2024classical}.
Hence, there have been significant interests and efforts in reducing the effect of photon loss and partial distinguishability, such as developing quantum error correction codes~\cite{michael2016new, chamberland2022building, marshall2022distillation, sivak2023real, faurby2024purifying}.

The key idea of most of the existing classical algorithms for simulating lossy systems is that their output state becomes classical when the loss rate increases as the system size grows~\cite{oszmaniec2018classical, garcia2019simulating, qi2020regimes, brod2020classical, oh2021classical, oh2023classical, liu2023complexity, oh2024classical, oh2025recent}.
In particular, Refs.~\cite{oszmaniec2018classical, garcia2019simulating, qi2020regimes, brod2020classical, oh2024classical} show that when the total transmission rate scales as $O(1/\sqrt{N})$, where $N$ is the input photon number, the circuit can be approximately described by a nonnegative quasi-probability distribution~\cite{mari2012positive, rahimi2016sufficient} and thus be classically simulated efficiently.
Also, other classical algorithms considering various types of noises and using an approximation of the output probability by low-degree polynomials from Refs.~\cite{kalai2014gaussian, renema2018classical, renema2018efficient, oh2023classical} assume Haar-random linear-optical circuits for their algorithms to be guaranteed to be efficient, which inevitably requires deep circuits~\cite{reck1994experimental} (See Supplemental Material~(SM)~\cite{supple} for more detailed discussion.).

Hence, another promising path to quantum advantage is to employ shallow-depth quantum circuits to minimize the photon-loss effect, where the loss rate does not necessarily increase with system size and the existing methods do not directly apply.
In fact, a worst-case constant-depth linear-optical circuit with single photons is proven hard for classical computers to exactly simulate unless the polynomial hierarchy~(PH) collapses to a finite level~\cite{brod2015complexity}.
Furthermore, there have been many attempts to prove the average-case hardness of approximate simulation of shallow-depth boson sampling circuits~\cite{van2021boson, go2024exploring, go2024computational}.
In addition, there have been proposals and proof-of-principle experiments that implement linear-optical circuits with high-degree connectivity, or even all-to-all connectivity~\cite{crespi2016suppression, imany2020probing, hu2020realization, lau2012proposal, shen2014scalable, chen2023scalable}.
Since one main reason to utilize shallow-depth circuits is to minimize the effect of noise and loss, a pertinent question that must be answered is whether shallow-depth circuits, even under the effect, are hard to classically simulate or it again destroys the potential quantum advantage.
It is worth emphasizing that since lossy boson sampling is hard to {\it exactly} classically simulate unless the PH collapses to a finite level (see the main text for more details.), an appropriate notion of simulation here is not an {\it exact} but {\it approximate} simulation, which is also a usual notion of quantum advantage~\cite{aaronson2011computational}.

To address this question, in this work, we analyze the computational complexity of constant-depth linear-optical circuits with all-to-all connectivity under photon loss and partial distinguishability and prove that when the input state is single photons, there exists a threshold of noise rates above which we can approximately simulate the system efficiently using classical computers.
The main idea is to associate a linear-optical circuit with a bipartite graph in such a way that the single photons correspond to one part of the vertices of the graph and the output modes correspond to the other part of the vertices of the graph and they are connected by edges if the single photons can propagate to the output modes through the linear-optical circuit.
We then appropriately adapt a well-known result of percolation theory from the study of network~\cite{shante1971introduction, sahimi1994applications} to bipartite graphs, which states that if some of the vertices of a graph of bounded degree are randomly removed, the resultant graph is divided into disjoint logarithmically-small-size graphs with high probability.
We then show that the effect of photon loss or partial distinguishability noise exactly corresponds to removing some of the vertices; thus, photon loss or partial distinguishability of photons effectively transforms constant-depth linear-optical circuits into logarithmically-small-size independent linear-optical circuits.
Consequently, we can simulate the entire circuit by individually simulating each small-size linear-optical circuit.
Our result suggests that while shallow-depth circuits are often believed to be less subject to loss and noise, it may not always be true because the entanglement constituted by shallow-depth circuits may be more easily destroyed by loss and noise.
Finally, we numerically analyze the effect for various architectures and discuss a general condition for our result to hold.

\section{Results}
\subsection{Linear-optical circuits with single photons}
Consider $M$-mode linear-optical circuits with $N$ single-photon input and arbitrary local measurement.
Linear-optical circuits consist of beam-splitter layers, which may be geometrically non-local, i.e., all-to-all connectivity.
This setup is the basis of boson sampling~\cite{aaronson2011computational} or the KLM protocol~\cite{knill2001scheme}.
While the latter typically require deep linear-optical circuits, we mainly focus on constant-depth circuits which are expected to be less subject to loss and noise.
Here, we define depth as the number of layers of beam splitters that can be implemented in parallel.
We emphasize that constant-depth boson sampling is proven hard to {\it exactly} classically simulate unless the PH collapses to a finite level~\cite{brod2015complexity}; thus, shallow-depth linear-optical circuits may be sufficient for quantum advantage.
Note, however, that it is not yet resolved whether constant-depth linear-optical circuits can yield quantum advantage for {\it approximate} simulation and that, as mentioned in the introduction, {\it approximate} simulation is the usual notation for claiming quantum advantage (e.g., Ref.~\cite{aaronson2011computational}).

\begin{figure}[t]
\includegraphics[width=240px]{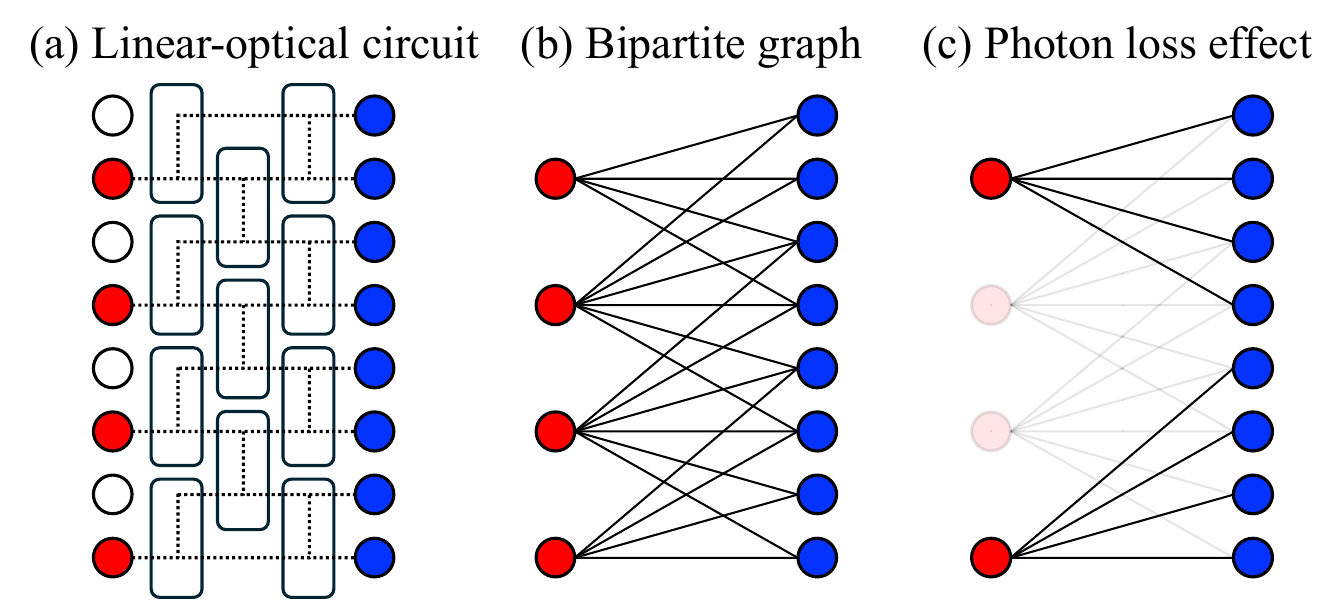}
\caption{(a) Linear-optical circuit with input photons on red dots and vacuum on empty dots on the left and output modes on blue dots on the right. Input and output modes are coupled by beam-splitter arrays with depth $d=3$, which make the input photons propagate through the dotted lines. (b) The corresponding bipartite graph. (c) When the second and third photons get lost, the corresponding vertices are removed from the graph. Consequently, the resultant bipartite graph is divided into two disconnected bipartite graphs.}
\label{fig:loss}
\end{figure}

Let us associate a linear-optical circuit with a bipartite graph (see Fig.~\ref{fig:loss}~(a)).
To do that, let $A$ and $B$ be the set of input modes that are initialized by single photons and the set of output modes to which the input photons can propagate through the given linear-optical circuit, respectively. 
Thus, $|A|=N$, and $|B|$ depends on the architecture and the circuit depth.
We then introduce a bipartite graph $G=(A,B,E)$ constituted by $A$ and $B$ for vertices on the left and right, respectively, and edges $E\subset A\times B$ between them.
The edges of the bipartite graph are determined by the light cone of input photons through the linear-optical circuit; namely, if a photon from an input mode corresponding to $v\in A$ can propagate to an output mode corresponding to $w \in B$, then the graph has an edge between the vertices, i.e., $(v,w)\in E$.
Here, we define the size of a bipartite as the number of vertices on the left-hand side, i.e., $|G|=|A|$.

Let us define $\Delta$ to be the maximum degree of the bipartite graph $G$, which is the maximum number of edges connected to a single vertex,
\begin{align}
    \Delta\equiv \max\big\{\max_{v\in A}|\{w\in B|(v,w)\in E\}|, \nonumber \\ 
    \max_{w\in B}|\{v\in A|(v,w)\in E\}|\big\}.
\end{align}
We note that when the depth of a linear-optical circuit is~$d$, the maximum degree of the associated bipartite graph is limited by $\Delta\leq 2^d$ because each beam splitter has two-mode input and output.
Hence, when the circuit depth is a constant, i.e., $d=O(1)$, the number of output modes relevant to each single photon input is also $O(1)$ for an arbitrary architecture.
When the circuit is geometrically local, say $D$-dimensional system, then each single photon can propagate up to $\Delta=O(d^D)$.
For example, when $D=1$, $\Delta\leq 2d+1$ and when $D=2$, $\Delta\leq 2d^2+2d+1$.

Using the introduced relation between linear-optical circuits and bipartite graphs and percolation theory, we investigate the complexity of simulating the linear-optical circuits under the effect of photon loss or partial distinguishability noise.

\subsection{Bipartite-graph percolation}
Percolation theory describes the behavior of graphs when adding or deleting vertices or edges of graphs~\cite{shante1971introduction, sahimi1994applications}.
It has been used in quantum information theory to study entanglement in quantum networks~\cite{acin2007entanglement, cuquet2009entanglement, pant2019percolation} and classical simulation of noisy quantum circuits~\cite{aharonov2000quantum, skinner2019measurement, trivedi2022transitions, rajakumar2024polynomial}, including a very recent result that presents an efficient classical algorithm simulating constant-depth noisy IQP circuits~\cite{rajakumar2024polynomial}.
Using a similar technique, we show that constant-depth linear-optical circuits with single photons are easy to classically simulate when the loss or noise rate is sufficiently high.
The key idea for this, together with the percolation lemma below, is that if a single photon is lost or becomes distinguishable from others, the system essentially loses interference induced by the photon.
This effect, from a graph perspective, is to remove or decouple the vertex from the graph for loss and distinguishability noise, respectively~(see below for more details).
To use this property, we adapt a result of percolation theory from Refs.~\cite{grimmett1999some, krivelevich2014phase, rajakumar2024polynomial} to bipartite graphs~(see Methods):
\begin{lemma}\label{lemma}
    Let $G=(A,B,E)$ be a bipartite graph of maximum degree $\Delta$.
    If we independently remove each $v\in A$ with probability $1-\eta$ with $\eta<1/\Delta^2$ and all edges incident to $v$ from $E$, the resultant bipartite graph is divided into $m$ bipartite graphs $\{G_i\}_{i=1}^m$ disconnected to each other and 
    \begin{align}
        \Pr(\max_i |G_i|>y)\leq N e^{-y(1-\eta\Delta^2-\log\eta\Delta^2)}.
    \end{align}
    Hence, when $\eta<1/\Delta^2$, with high probability $1-\epsilon$, the largest graph size in $\{G_i\}_{i=1}^m$ is smaller than or equal to
    \begin{align}
        y^*=\frac{\log(N/\epsilon)}{1-\eta\Delta^2-\log(\eta\Delta^2)}=O(\log(N/\epsilon)).
    \end{align}
\end{lemma}
From the physical perspective, it implies that for linear-optical circuits with maximum degree~$\Delta$, losing~${1-1/\Delta^2}$ portion of input photons is so significant that the remaining system can be effectively described by independent small-size systems.

\subsection{Photon loss effect}
Now, let us consider photon loss on input photons and observe the correspondence between photon loss and the removal of some vertices in the associated bipartite graph as in Lemma~\ref{lemma}.
When a single photon is subject to a loss channel with loss rate $1-\eta$, it transforms as
\begin{align}
    |1\rangle\langle 1|\to (1-\eta)|0\rangle\langle 0|+\eta|1\rangle\langle1|.
\end{align}
Then, while there is no effect with probability $\eta$, the vertex is removed from $A$ with probability $1-\eta$.
Consequently, when $N$ input single photons are subject to a loss channel with loss rate $1-\eta$, the effect is to remove each vertex with probability $1-\eta$ independently, which is exactly the same procedure in Lemma~\ref{lemma}~(see Fig.~\ref{fig:loss} for the illustration).

By taking advantage of this relation and Lemma~\ref{lemma}, we propose a classical algorithm simulating linear-optical systems with $N$ single photon input when $\eta\Delta^2<1$.
First, we remove each vertex in $A$ with probability $1-\eta$ for the associated bipartite graph with a given linear-optical circuit with single photons.
For the resultant graph, we identify components $\{G_i\}_{i=1}^m$ that are disconnected from each other.
If the size of any connected component $G_i$ obtained from the first step is larger than $y^*$, then we return to the first step.
Otherwise, we now have $\{G_i\}_{i=1}^m$ with $|G_i|\leq y^*=O(\log (N/\epsilon))$ for all $i$'s.
Since the number of photons for each connected component scales logarithmically, we can expect that it is easy to classically simulate.
Here, $\epsilon$ is chosen to be the desired total variation distance (TVD) (see below).

More specifically, since the input photon number in each $G_i$ is at most $y^*$ and the number of relevant output modes is at most $\Delta y^*$, the associated Hilbert space's dimension is upper-bounded by
\begin{align}
    \binom{\Delta y^*+y^*-1}{y^*}
    \leq [e(\Delta+1)]^{y^*}=\poly(N/\epsilon),
\end{align}
where we used $\binom{n}{k}\leq (\frac{ne}{k})^k$.
Thus, when $\Delta=O(1)$, like constant-depth circuits, the dimension is upper-bounded by $\poly(N/\epsilon)$.
Therefore, since writing down the output state and the relevant operators takes polynomial time in $N/\epsilon$ for any local measurement, we can efficiently simulate the system.
For photon-number detection, like boson sampling~\cite{aaronson2011computational}, one may simply use the Clifford-Clifford algorithm whose complexity is given by $\tilde{O}(2^{y^*})=\poly(N/\epsilon)$~\cite{clifford2018classical}.

If $\Delta$ scales with the system size $N$, i.e. beyond constant depth, and the measurement is not photon-number detection, the above counting gives us a superpolynomially increasing dimension in $N/\epsilon$.
For this case, to be more efficient, consider a $y^*$ number of single photons as an input and note that a linear-optical circuit~$\hat{U}$ transforms the creation operators of input modes $\hat{a}^\dagger_j$ as
\begin{align}
    \hat{a}_j^\dagger 
    \to \sum_{k=1}^{M^*} U_{jk} \hat{a}_k^\dagger
    =\sum_{k=1}^L U_{jk} \hat{a}_k^\dagger+\sum_{k=L+1}^{M^*} U_{jk} \hat{a}_k^\dagger
    \equiv \hat{B}^{L,\dagger}_{\text{u},j}+\hat{B}^{L,\dagger}_{\text{d},j},
\end{align}
where $U$ is the ${M^*}\times {M^*}$ unitary matrix characterizing the linear-optical circuit with $M^*$ being the relevant number of modes, and we set a bipartite between output modes $[1,\dots,L]$ and $[(L+1),\dots,{M^*}]$.
Then, the output state can be written as
\begin{align}
    |\psi_\text{out}\rangle
    =\prod_{j=1}^{y^*} \left( \hat{B}^{L,\dagger}_{\text{u},j}+\hat{B}^{L,\dagger}_{\text{d},j}\right)|0\rangle
    \equiv \sum_{\bm{x}\in\{\text{u},\text{d}\}^{y^*}}\prod_{j=1}^{y^*}\hat{B}_{x_j,j}^{L,\dagger}|0\rangle,
\end{align}
Therefore, for any bipartition, the output state can be described by at most $2^{y^*}=\poly(N/\epsilon)$ singular values, implying that a matrix product state~(MPS) with bond dimension $\poly(N/\epsilon)$ can describe the state~\cite{vidal2003efficient}, which can be found by time-evolution block decimation~\cite{schollwock2011density}.
Note that if the circuit is not linear-optical, the above decomposition is no longer valid because output bosonic operators may contain a product of two operators from each partition (e.g., $\hat{a}_1^\dagger \hat{a}_M^\dagger$).

The remaining question is the algorithm's error caused by skipping the case where $\max_{i}|G_i|>y^*$.
Then, the output probability $q$ of such an algorithm is given by
\begin{align}
    q(\bm{m})=p(\bm{m}|E),
\end{align}
where $p(\bm{m}|E)$ is the conditional probability on the case~$E$ where $\max_{i}|G_i|\leq y^*$; hence, the true output probability is written as
\begin{align}
    p(\bm{m})=p(\bm{m}|E)p(E)+p(\bm{m}|E^\perp)p(E^\perp),
\end{align}
where $E^\perp$ is the case where $\max_{i}|G_i|>y^*$.
Then, the TVD between $q(\bm{m})$ and $p(\bm{m})$ can be shown smaller than~$\epsilon$~(see Methods).
Finally, an overhead per sample from the restart is given by $1/p(E)\leq 1/(1-\epsilon)=O(1)$.
Thus, we have
\begin{theorem}
    For a given loss rate $1-\eta$ and a linear-optical circuit of maximum degree $\Delta$ with $N$ single-photon input, if $\eta <1/\Delta^2$, there exists a classical algorithm that can approximately simulate the circuit in $\poly(N,1/\epsilon)$ within TVD $\epsilon$.
\end{theorem}
Hence, for constant-depth linear-optical circuits, $d=O(1)$ and $\Delta=O(1)$, there is a threshold of loss rate above which it becomes classically easy to simulate.
Note that the actual threshold depends on the architecture~(see below for further discussion).
We emphasize that the measurement basis can be arbitrary as long as it is local; thus, our theorem can be employed beyond boson sampling.

Here, the notion of approximate simulation is crucial because the exact classical simulation of constant-depth {\it lossy} linear-optical circuits is hard unless the PH collapses to a finite level.
This can easily be shown by noting that postselecting no loss case of lossy boson sampling is equivalent to lossless boson sampling and constant-depth boson sampling with post-selection is post-BQP due to the measurement-based quantum computing~\cite{brod2015complexity}.
Thus, if lossy constant-depth boson sampling can be exactly simulated using classical algorithms efficiently, $\text{PH}\subset \text{P}^\text{PP}=\text{P}^\text{post-BQP}=
\text{P}^\text{post-BPP}$~\cite{aaronson2005quantum}, which contradicts the fact that $\text{P}^\text{post-BPP}$ is in the
PH~\cite{han1997threshold} assuming that the PH is infinite.
Hence, the complexity of exact simulation and approximate simulation of constant-depth lossy boson sampling has a gap when the loss rate is above the threshold.

We also consider cases where each layer of beam splitters has transmission rate $\eta_1<1$, thus $\eta=\eta_1^d$.
Since $\Delta\leq 2^d$ for any architecture and loss channel with a uniform loss rate commutes with beam splitters~(see, e.g., Refs.~\cite{garcia2019simulating, oh2021classical}), we have the following corollary:
\begin{corollary}
    For a linear-optical circuit with $N$ single-photon input and a transmission rate per layer $\eta_1$, if $\eta_1<1/4$, there exists a classical algorithm that can approximately simulate the circuit in $\poly(N,1/\epsilon)$ within TVD $\epsilon$.
\end{corollary}
For this case, the presented MPS method is crucial because the depth may not be constant.

\subsection{Partial distinguishability noise}
A similar observation can be used when input photons are partially distinguishable, which is another important noise model in optical systems~\cite{tichy2015sampling, renema2018efficient, renema2018classical, moylett2019classically}.
The underlying physical mechanism that causes photons to be partially distinguishable is other degrees of freedom of photons, such as polarization and temporal shapes.
Consequently, when the other degrees of freedom do not match perfectly, the overlap of the wave functions of a pair of photons becomes less than 1 (see Ref.~\cite{tichy2015sampling} for more discussion.).
For simplicity, we assume that the overlap for any pairs of photons is uniform as $0<x<1$.

\begin{figure}[t!]
\includegraphics[width=240px]{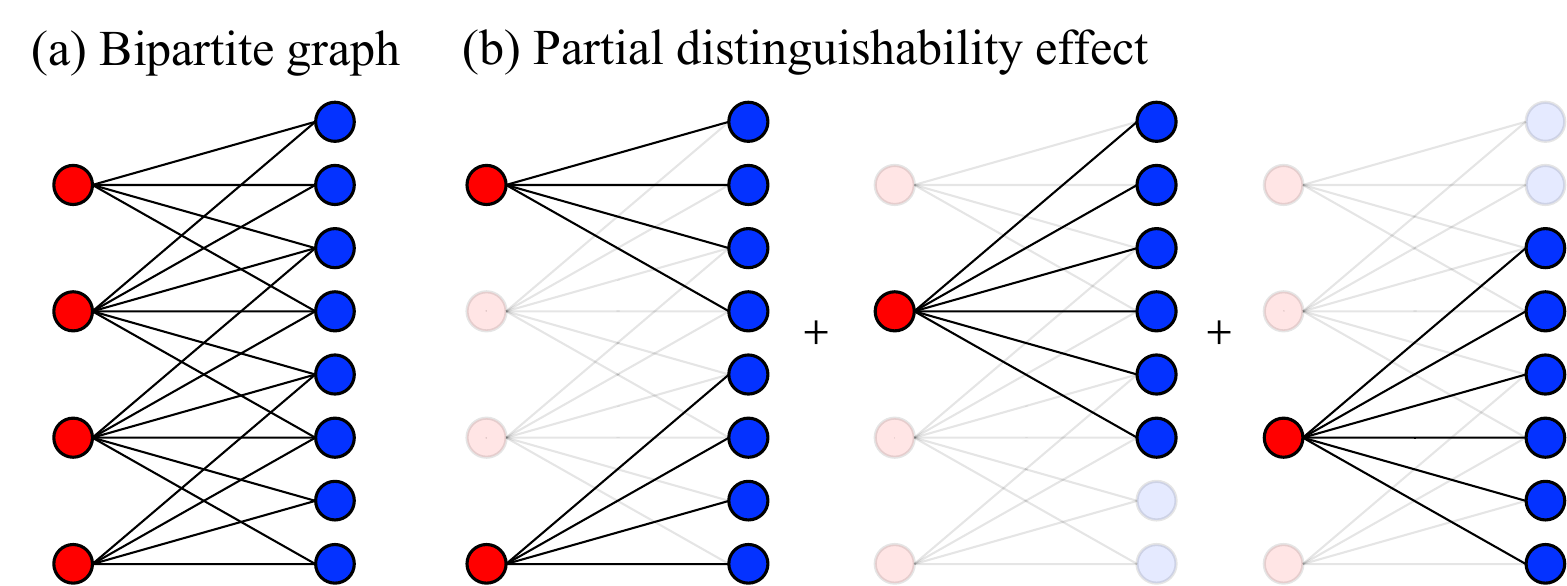}
\caption{Effect of partial distinguishability. (a) Bipartite graph from Fig.~\ref{fig:loss}. (b) When the second and third photons become distinguishable from other photons, they are isolated from the other vertices. Therefore, the resultant system can be simulated by simulating individual smaller systems.}
\label{fig:dist}
\end{figure}

Ref.~\cite{moylett2019classically} shows that such a model transforms an $N$ single-photon state to the following density matrix
\begin{align}
    \hat{\rho}=\sum_{k=0}^N p_k \sum_{I\subset [N],|I|=k}\hat{\rho}_I,
\end{align}
where $\hat{\rho}_I$ is the state whose $I$ elements are indistinguishable and others are distinguishable and $p_k\equiv x^k(1-x)^{N-k}$.
Then, the quantum state of $N$ partially distinguishable single photons is equivalent to the mixture of an $N$-particle state obtained by randomly selecting $k$ particles following a binomial distribution with success probability $x$ and setting them indistinguishable bosons and other fully distinguishable particles.
Therefore, the remaining $N-k$ photons do not interfere with others.
From the graph perspective, it corresponds to decoupling the corresponding vertices from the original graph as illustrated in Fig.~\ref{fig:dist}.
A difference from photon loss is that we construct bipartite graphs with each removed vertex.
Lemma~\ref{lemma} still applies here because the resultant bipartite graphs still have the maximum size $O(\log (N/\epsilon))$ with high probability.
Thus, 
\begin{theorem}
    For partial distinguishability $1-x$ and a linear-optical circuit of maximum degree $\Delta$ with $N$ single-photon input, if $x <1/\Delta^2$, there exists a classical algorithm that can approximately simulate the circuit in $\poly(N,1/\epsilon)$ within TVD $\epsilon$.
\end{theorem}

\begin{figure}[t!]
\includegraphics[width=240px]{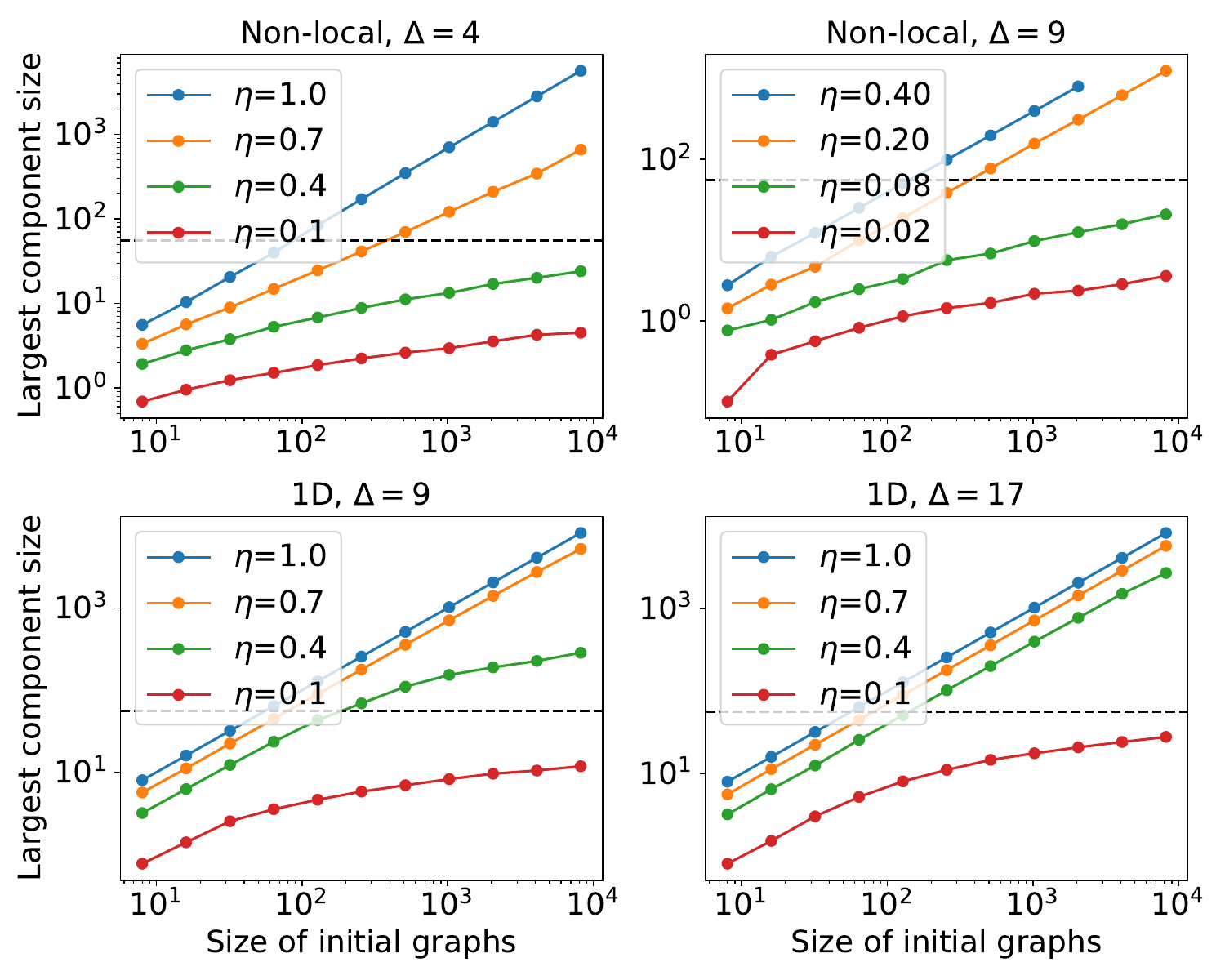}
\caption{Largest size of the component obtained by removing vertices in $A$ of initial size $N$ with probability $1-\eta$. The dashed lines indicate the size of graphs with 56 vertices, corresponding to the largest permanent computed so far using a supercomputer~\cite{elbek2025superman}, i.e., the boundary of our classical algorithm. (See the main text for the details and Supplemental Material~\cite{supple} for different scales of each axis.)}
\label{fig:numerical}
\end{figure}

\subsection{Numerical result}
To clearly see the percolation effect, we numerically investigate the maximum size of components $\max_i |G_i|$ after randomly removing vertices from $A$.
We consider two cases: linear-optical circuits (i) with non-local beam splitters and (ii) with 1D-local beam splitters.
For simplicity of the numerical simulation, in the non-local case, for a given input photon number~$N$ and the number of modes $M=8N$, we randomly generate~$\Delta$ edges from each input, and, in the 1D case, we generate $\Delta$ edges from each input mode to the~$\Delta$ closest output modes.

For a fixed $\Delta$ and different loss rates, we increase the number of input modes $N$ and analyze the largest component size, $\max_i |G_i|$, which is showcased in Fig.~\ref{fig:numerical}.
Clearly, the largest component size increases in distinct ways depending on the loss rates; when the loss rate is sufficiently low, it increases linearly, while the loss rate is high enough, it starts to increase logarithmically as expected by Lemma~\ref{lemma}~(See SM~Sec.~S2~\cite{supple}.).
We also emphasize that the transition point depends on the architecture.
For instance, when $\Delta=9$, the transition occurs between ${\eta=0.4}$ and ${\eta=0.7}$ for 1D systems, whereas it occurs between ${\eta=0.02}$ and ${\eta=0.14}$, implying that the non-local structure is much more robust than the 1D structure.
Therefore, while Lemma~\ref{lemma} presents the worst-case threshold, the actual threshold may depend on the details of the architecture, such as the geometry of the architecture~(e.g., non-local beam splitters or geometrically local beam splitters.) and how many output modes are relevant.

Finally, we note that the state-of-the-art (Gaussian) boson sampling experiments suffer from a loss rate around $0.5-0.7$, i.e., $\eta\approx 0.3-0.5$~\cite{zhong2020quantum, zhong2021phase, madsen2022quantum, deng2023gaussian} and that the largest single-photon boson sampling so far, to the best of our knowledge, is up to 20 photons~\cite{wang2019boson}.
Figure~\ref{fig:numerical} shows that while a very shallow-depth nonlocal circuit with $\Delta=4$ is easy to simulate using our algorithm around this regime even when the initial number of photons is as large as $10^4$, as the circuit depth increases (e.g., $\Delta=9$), our algorithm starts to require a computational cost beyond the currently available supercomputers.
A similar behavior can be observed for 1D linear-optical circuits.
Overall, unless the loss rate is very low, around an initial hundred photons are required for shallow-depth boson sampling to be beyond our classical algorithm.
It is worth emphasizing that this boundary does not necessarily imply the boundary of classical algorithms because there may exist better classical algorithms than the presented one.

\subsection{Discussion on more general cases}
We now discuss the possibility of generalizing the above results to more general setups.
First of all, for photon-loss cases, it is not hard to see that even if we replace single photons with general Fock states $|n\rangle$, a similar theorem holds.
This is because a Fock state with photon number $n$ transforms under photon loss as
\begin{align}
    |n\rangle\langle n|
    &\to \sum_{k=0}^n \binom{n}{k} \eta^k (1-\eta)^{n-k}|k\rangle\langle k| \\ 
    &=(1-\eta)^n|0\rangle\langle 0|+\sum_{k=1}^n \eta^k (1-\eta)^{n-k}|k\rangle\langle k|.
\end{align}
Therefore, we can follow the same procedure as single-photon cases.
A difference is that we now sample from a binomial distribution with failure probability $(1-\eta)^n$, which corresponds to vacuum input, and thus, the associated vertex is removed from the bipartite graph.
Consequently, the percolation threshold becomes ${[1-(1-\eta)^n]\Delta^2 <1}$.
For this case, since each input mode has at most $n$ photons, the associated Hilbert space's dimension for $y^*$ is at most 
\begin{align}
    \binom{\Delta y^*+n y^*-1}{ny^*}=\poly(N/\epsilon).
\end{align}
Similarly, for more general input states than Fock states, if the largest photon number of each input state for each mode is constant, the Hilbert space's dimension is at most polynomial in $N/\epsilon$; more generally, as long as the largest total photon number is bounded by a linear function of $y^*$, the Hilbert space's dimension is still upper-bounded by $\poly(N/\epsilon)$.
For larger depth $d=\omega(1)$, we again need to use the MPS method~(see Methods).

Hence, a sufficient condition for the percolation result to apply is that the lossy input state is written as $\hat{\rho}=(1-p)|0\rangle\langle 0|+p\hat{\sigma}$, where $0<p<1$, $\hat{\sigma}\geq 0$, $\Tr[\hat{\sigma}]=1$, and $\hat{\sigma}$ can be written in the Fock basis with at most a constant photon number.
As the Fock-state example implies, the threshold value depends on how the input state transforms under a loss channel.
Thus, an analytic way to compute the threshold value for arbitrary input states has to be further studied.
Also, we emphasize that the assumption that the circuit is linear-optical is important because otherwise, we may be able to apply an operation that generates photons in the middle of the circuit, such as a squeezing operation, after the loss channel on the input.

\section{Discussion}
We showed that a threshold of loss or noise rate exists above which classical computers can efficiently simulate constant-depth linear-optical circuits with certain input states.
Thus, shallow-depth circuits may also be vulnerable to loss and noise because the entanglement in the system constructed by shallow-depth circuits may be more easily annihilated by noise.



An interesting future work is to find a general condition for input states under which the percolation result gives the easiness result.
Whereas Fock states' threshold is easily found, it is not immediately clear to analytically find the threshold of more general quantum states, such as Gaussian states~\cite{serafini2017quantum}.
Also, while our results hold for arbitrary architecture as long as the depth is constant, the depth limit might be pushed further depending on the details of the architecture, such as the geometry of the circuits and the input state configuration~\cite{deshpande2018dynamical, oh2022classical, qi2022efficient}.
Conversely, investigating the possibility of the hardness of constant-depth boson sampling with a loss rate below the threshold is another interesting future work enabling us to demonstrate quantum advantage even under practical loss and noise effects.
Finally, we can easily see that the percolation lemma immediately applies for a continuous-variable erasure channel considered in Refs.~\cite{wittmann2008quantum, lassen2010quantum, zhong2023information}.
Thus, we may be able to apply a similar technique to other noise models, such as more general Gaussian noise~\cite{serafini2017quantum}.

\section*{Methods}
\subsection{Proof of Lemma~1}\label{app:proof}
We provide the proof of Lemma~1 in the main text.
The proof is based on Refs.~\cite{grimmett1999some, krivelevich2014phase, rajakumar2024polynomial} and is adapted to bipartite-graph cases.
\begin{proof}
Suppose we have an $N$ by $M$ bipartite graph $G=(A,B,E)$ with maximum degree $\Delta$ and then remove some vertices on the left-hand side $A$ with probability $1-\eta$ and edges connected to them.

We present an algorithm that constructs a random graph $G'=(A',B',E')$ as described in the lemma.
Denote the set of the vertices on the left-hand side as $A$ and on the right-hand side as $B$.
We initialize this to be the empty graph.
Then, we construct $S$ by querying the vertices on $A$.
A query to vertex $v\in A$ succeeds with probability $\eta$, in which case the vertex is added to $S$.
When $S=\emptyset$, the algorithm initializes $S$ by querying all unqueried vertices in $G$ until the first successful query.
When $S\neq \emptyset$, the algorithm queries all unqueried vertices in $N_G(S)$, where $N_G(S)$ is a subset of $A$ which is connected by $S$ through $B$ by one step.
The size of $N_G(S)$ is upper-bounded by $|S|\Delta^2$.
Whenever the algorithm runs out of unqueried vertices, it adds $S$, the vertices in $B$ connected to $S$, and corresponding edges to $G'$ and resets $S$ and continues.
The algorithm finishes when there are no more unqueried vertices in $A$.
Note that when the algorithm adds $S$ and its corresponding vertices in $B$ and edges to $G'$, the added graph is always disconnected to the one in every step, which results in the set of disjoint components $\{G_i\}_{i=1}^m$, where $m$ is the number of steps.

If there is a component $G_i$ of size $y+1$ or higher, $|S|$ must have reached $y+1$ at some point.
At this point, suppose the most recent vertex added to $S$ is labeled $v$.
To reach this point, we could have made at most $|S\cup N_G(S-v)|\leq \Delta^2(|S|-1)=y\Delta^2$ queries with exactly $y+1$ being successful.
Hence, the probability of forming a $G_i$ of size $y+1$ or higher is upper bounded as
\begin{align}
    \Pr(|G_i|>y)\leq \Pr(\text{Bin}(y\Delta^2,\eta)>y).
\end{align}
Here, $\Pr(\text{Bin}(y\Delta^2,q)>y)$ means the probability of obtaining more than $y$ successes from binomial sampling out of $y\Delta^2$ trials with success probability $\eta$.
Using the Chernoff bound with mean $\mu=\eta y\Delta^2$,
\begin{align}
    \Pr(\text{Bin}(y\Delta^2,\eta)>(1+\delta)\mu)\leq \left(\frac{e^{-\delta}}{(1+\delta)^{1+\delta}}\right)^\mu,
\end{align}
with setting $1+\delta=1/(\Delta^2 \eta)$,
\begin{align}
    \Pr(|G_i|>y)
    &\leq \left(\frac{e^{1-1/\Delta^2 \eta}}{(1/\Delta^2 \eta)^{1/\Delta^2 \eta}}\right)^{\eta y\Delta^2}  \\
    &\leq \left(\frac{e^{\Delta^2 \eta-1}}{(1/\Delta^2 \eta)}\right)^{y} \\
    &\leq e^{-y(1-\eta \Delta^2-\log\eta\Delta^2)}.
\end{align}
By applying the union bound,
\begin{align}
    \Pr(\max_i|G_i|>y) 
    &\leq \sum_{i=1}^m \Pr(|G_i|>y)  \\
    &\leq Ne^{-y(1-\eta \Delta^2-\log\eta\Delta^2)}.
\end{align}
\end{proof}

It is worth emphasizing that the lemma can be understood from a previous result~\cite{krivelevich2014phase, rajakumar2024polynomial} by defining a graph based on the bipartite graph in such a way that the graph's vertices are $A$ and two vertices have an edge if they are connected by a vertex in $B$ in the original bipartite graph.
Physically, two input modes are connected if the photons from them can be outputted in the same output mode, i.e., they can interfere with each other.
Using the relation between the underlying bipartite and the new graph $G$, we can easily see that the maximum degree of the graph $G$ is upper-bounded by $\Delta^2$.

\subsection{Upper bound of total variation distance}
We derive the upper bound of the approximation error caused by skipping the case where the connected component's size is larger than $y^*$.
Recall that the output probability $q$ of such an algorithm is given by
\begin{align}
    q(\bm{m})=p(\bm{m}|E).
\end{align}
where $p(\bm{m}|E)$ is the true conditional probability of the case $E$ where the maximum size of the connected components is smaller than or equal to $y^*$; hence, the true probability is written as
\begin{align}
    p(\bm{m})=p(\bm{m}|E)p(E)+p(\bm{m}|E^\perp)p(E^\perp),
\end{align}
where $E^\perp$ is the case where the maximum size of the connected components is larger than $y^*$.
Then, the total variation distance between the suggested algorithm's output probability $q(\bm{m})$ and the true output probability $p(\bm{m})$ can be shown to be smaller than $\epsilon$:
\begin{align}
    &\sum_{\bm{m}}|p(\bm{m})-q(\bm{m})| \\
    &=\sum_{\bm{m}}|p(\bm{m}|E)p(E)+p(\bm{m}|E^\perp)p(E^\perp)-p(\bm{m}|E)| \\
    &\leq \sum_{\bm{m}}|p(\bm{m}|E)p(E)-p(\bm{m}|E)|+\sum_{\bm{m}}p(\bm{m}|E^\perp)p(E^\perp) \\
    &=\sum_{\bm{m}}p(\bm{m}|E)|p(E)-1|+p(E^\perp) \\
    &=2p(E^\perp) \leq 2\epsilon.
\end{align}

\subsection{Matrix product state for more general states than single photons}\label{app:MPS}
In this section, we show that any linear-optical circuits with $N$ input states that have a constant maximum photon number can be simulated by matrix product state with bond dimension at most $c^N$; thus, the computational cost is exponential in $N$ with a constant~$c$.

Let us consider a linear-optical circuit $\hat{U}$, which transforms the creation operators of input modes $\hat{a}^\dagger_j$ into the creation operators of output modes $\hat{b}^\dagger_j$ as
\begin{align}
    \hat{a}_j^\dagger \to 
    \sum_{k=1}^M U_{jk} \hat{a}_k^\dagger
    =\sum_{k=1}^L U_{jk} \hat{a}_k^\dagger+\sum_{k=L+1}^M U_{jk} \hat{a}_k^\dagger
    \equiv \hat{B}^{L,\dagger}_{\text{u},j}+\hat{B}^{L,\dagger}_{\text{d},j},
\end{align}
where $U$ is an $M\times M$ unitary matrix characterizing the linear-optical circuit $\hat{U}$.
When we prepare $N$ input states with maximum photon number $n_\text{max}$,
\begin{align}
    \sum_{n=0}^{n_\text{max}}c_n|n\rangle
    =\sum_{n=0}^{n_\text{max}}\frac{c_n}{\sqrt{n!}} \hat{a}^{\dagger n}|0\rangle,
\end{align}
the total input state transforms as
\begin{align}
    |\psi_\text{in}\rangle
    &=\prod_{j=1}^N\left(\sum_{n=0}^{n_\text{max}}\frac{c_n}{\sqrt{n!}} \hat{a}_j^{\dagger n}\right)|0\rangle  \\
    &\to \prod_{j=1}^N\left[\sum_{n=0}^{n_\text{max}}\frac{c_n}{\sqrt{n!}} \left(\hat{B}_{\text{u},j}^{L,\dagger}+\hat{B}_{\text{d},j}^{L,\dagger}\right)^{n}\right]|0\rangle \\
    &=\prod_{j=1}^N\left[\sum_{n=0}^{n_\text{max}}\frac{c_n}{\sqrt{n!}}\sum_{k=0}^{n}\binom{n}{k}\left(\hat{B}_{\text{u},j}^{L,\dagger}\right)^k\left(\hat{B}_{\text{d},j}^{L,\dagger}\right)^{n-k}\right]|0\rangle.
\end{align}
Thus, the output state can be written as the linear combination of at most $[(n_\text{max}+1)(n_\text{max}+2)/2]^N$ vectors that are a product of a vector in $\text{u}$ and a vector in $\text{d}$.
Therefore, as long as $n_\text{max}$ is constant, the output state requires at most an exponential number of $N$ singular values; hence, the matrix product state method can simulate the system.

\begin{acknowledgements}
We thank Byeongseon Go and Senrui Chen for interesting and fruitful discussions.
This research was supported by the National Research Foundation of Korea (NRF) Grants (No. RS-2024-00431768 and No. RS-2025-00515456) funded by the Korean government (Ministry of Science and ICT~(MSIT)).
\end{acknowledgements}

\section*{Author contributions}
C.O. developed the theory, implemented the numerical experiments, and wrote the paper.

\section*{Competing interests}
The author declares no competing interests.

\bibliography{reference.bib}

\newpage

\end{document}


\definecolor{navy}{RGB}{46,72,102}
\definecolor{pink}{RGB}{219,48,122}
\definecolor{grey}{RGB}{184,184,184}
\definecolor{yellow}{RGB}{255,192,0}
\definecolor{grey1}{RGB}{217,217,217}
\definecolor{grey2}{RGB}{166,166,166}
\definecolor{grey3}{RGB}{89,89,89}
\definecolor{red}{RGB}{255,0,0}

\preprint{APS/123-QED}

\title{Supplemental Material for ``Classical simulability of constant-depth linear-optical circuits with noise"}
\author{Changhun Oh}
\email{changhun0218@gmail.com}
\affiliation{Department of Physics, Korea Advanced Institute of Science and Technology, Daejeon 34141, Korea}

\maketitle

\section{Comparison to related results}\label{app:literature}
In this section, we provide a detailed comparison of the main results of this paper to related results in the literature.
Since understanding the complexity of boson sampling in practice is crucial for demonstrating quantum advantage using experimental boson sampling, there are a plethora of related studies.
We review the related results and highlight the distinction between our main and existing results.
Here, we will denote $M$ and $N$ as the number of modes and photons, respectively, and assume that $M=\poly(N)$ as usual.

\subsection{Our main result}
To compare with the previous results, let us summarize the main result of the presented work and emphasize the problem setup.
In this work, we focus on constant-depth linear-optical circuits with a constant rate of loss or noise and prove that there is a noise threshold above which there is a classical algorithm that can simulate the system.
We emphasize that our circuits can be arbitrary as long as they are linear-optical and constant-depth.
In particular, it may be on an arbitrary architecture, such as it may have all-to-all connectivity, and thus does not assume geometrically local architecture.
Also, our classical simulability is for the worst-case linear-optical circuits, i.e., the classical algorithm works for any linear-optical circuits if the noise rate is high enough, which is in stark contrast to many of the previous results which assume linear-optical circuits with Haar-random unitary matrices and whose classical algorithms are efficient in average case, not in the worst case.
While such a Haar-randomness is crucial for their analysis, since a Haar-random linear-optical circuit requires beyond constant-depth circuits~\cite{reck1994experimental}, the regimes of interest do not overlap.
Finally, many of the previous results do not generalize to beyond photon-number basis measurement because they exploit the properties of permanent, our classical algorithm does not assume photon-number basis measurement.




\subsection{Classical simulations based on quasi-probability distribution}
Let us now directly compare our results to existing results.
Ref.~\cite{rahimi2016sufficient} presents a way to understand how loss and dark count affect the classical simulability of boson sampling using quasi-probability distribution based on Ref.~\cite{mari2012positive}.
The main idea is that when a given quantum circuit can be represented by non-negative quasi-probability distribution, such as the Wigner function or its variants, the quantum circuit can be efficiently simulated using a classical algorithm.
They then observe that when there is an effect of loss and dark count, the relevant Fock-state boson sampling circuit can be represented by nonnegative quasi-probability distribution, unlike its noiseless counterpart.
A caveat is that the classical simulability presented in this work has to assume a large amount of dark count; thus it does not account for the effect of photon loss solely, which is refined in Ref.~\cite{garcia2019simulating} using approximation.

Ref.~\cite{garcia2019simulating} analyzes the effect of loss for two different regimes: (i) $d\geq O(\log M)$ and (ii) $d\leq O(\log M)$. 
While the second case is more relevant to our work because we focus on constant-depth boson sampling, it is still worth reviewing the first case as well.
For case (i), they consider a lossy boson sampling circuit in which a constant rate of loss occurs for each depth, i.e., the total transmission rate in the entire circuit is given by $\eta^{d}$, where $\eta$ is the transmission rate per depth. 
They then observe that uniform loss channels in the circuit can be assumed to occur only on the input single-photon states because they commute with beam splitters and that when the total transmission rate scales as $O(1/\sqrt{N})$, one may approximate the input lossy single-photon state to be a thermal state whose $P$-representation is nonnegative and simulate the lossy boson sampling circuit approximately using the classical algorithm from Ref.~\cite{rahimi2016sufficient}.
Note that the thermal-state approximation entails a small approximation error when the total transmission rate scales as $O(1/\sqrt{N})$, which is thus valid only when the depth is sufficiently large so that the total transmission rate decreases as we increase the system size.
In stark contrast, this is obviously not the case for constant-depth boson sampling because the total transmission rate $\eta^d$ in the latter is constant.
Hence, this method does not immediately apply to the constant-depth boson sampling we consider in the presented work.
We emphasize that many of the other previous results, whose details are provided below, also study the simulability when the total transmission rate decreases (equivalently, the total loss rate or total noise rate converges to 1.) and these methods do not immediately apply for constant-depth boson sampling because of this reason, which is our main focus.
Note that a similar idea applies to lossy Gaussian boson sampling circuits and its explicit classical algorithm and analysis are given in Ref.~\cite{qi2020regimes} and is further improved in Ref.~\cite{oh2024classical} to simulate the state-of-the-art Gaussian boson sampling experiments~\cite{zhong2020quantum, zhong2021phase, madsen2022quantum, deng2023gaussian}.

In fact, the second case (ii) deals with a shallow-depth boson sampling.
Their approach is to employ the matrix product state~(MPS) method~\cite{vidal2003efficient}, which is a powerful tool to simulate a system with a low entanglement.
However, we emphasize that the MPS method they present works efficiently only when the circuit architecture is one-dimensional and it no longer works if we consider an architecture larger than one-dimensional systems.
Notably, the method we use is valid for arbitrary architectures, which may have all-to-all connections.
This is crucial because many recent experiments are beyond one-dimensional (e.g., Refs.~\cite{zhong2020quantum, zhong2021phase,madsen2022quantum,deng2023gaussian}).
Therefore, the two classical algorithms presented in Ref.~\cite{garcia2019simulating} cannot simulate a constant-depth lossy boson sampling with an arbitrary architecture, which is our main focus.

Ref.~\cite{oszmaniec2018classical} also studies the effect of loss and presents another classical algorithm that simulates lossy boson sampling when the total transmission rate scales as $O(1/\sqrt{N})$, the main idea of which is to approximate the lossy quantum state by a separable state.
As mentioned above, this scaling appears when the circuit depth increases as the system size.
While Refs.~\cite{garcia2019simulating, oszmaniec2018classical} assume a uniform loss circuit, Ref.~\cite{brod2020classical} extends these classical algorithms to apply a nonuniform loss circuit.

\subsection{Classical simulations based on approximating output probabilities of noisy boson sampling}
On the other hand, unlike the above results that exploit the quasi-probability distribution, Ref.~\cite{kalai2014gaussian} directly studies how the output probability of boson sampling changes under noise.
Here, the noise they consider can be interpreted as the fluctuation of linear-optical circuits and is different from photon loss and partial indistinguishability.
We emphasize that Ref.~\cite{kalai2014gaussian} and its subsequent works~\cite{shchesnovich2019noise, renema2018efficient, renema2018classical, oh2023classical} discussed below assume Haar-random linear-optical circuits, which require a deep quantum circuit~\cite{reck1994experimental}.
Hence, their results do not overlap with the regime of interest in the presented work.

To be more specific, Ref.~\cite{kalai2014gaussian} observes that when the noise rate is sufficiently high, the output probability of noisy boson sampling can be efficiently approximated.
The main idea is that when the noise rate is high, the output probability can be approximated by a low-degree polynomial, which is easy to compute.
Refs.~\cite{shchesnovich2019noise, oh2023classical} extend this approximate computing of the output probability to a classical algorithm for sampling.~(Note that in many cases, approximation of the output probabilities is not sufficient for approximate simulation of sampling~\cite{pashayan2020estimation}.)
Using a similar observation and a similar low-degree polynomial approximation technique to Ref.~\cite{kalai2014gaussian}, Ref.~\cite{renema2018efficient} generalizes the idea to partial distinguishability of photons and its effect on classical simulability and Ref.~\cite{renema2018classical} studies the combined effect of partial distinguishability and photon loss.
As we emphasized above, all these results assume a Haar-random linear-optical circuit for their approximation error analysis whereas our result focuses on constant-depth linear-optical circuits which do not form a Haar-random linear-optical circuit.
Hence, the previous results and ours do not overlap and are thus rather complimentary to each other.
Note that this idea has been generalized to Gaussian boson sampling~\cite{renema2020simulability}.




\subsection{Classical algorithms for shallow-depth boson sampling}
Meanwhile, there are other related works to the presented work that study the classical simulability of shallow-depth boson sampling.
We first emphasize that while the works mentioned below study the classical simulability of shallow-depth boson sampling without introducing the effects of noise or loss, our work focuses on the combined effects, i.e., the effects of the noise and loss on shallow-depth boson sampling, which is an important distinction from the previous results.
Also, the combination of the shallowness and the effects of loss or noise is the key to our result.

As mentioned above, Ref.~\cite{garcia2019simulating} introduces the MPS for simulating 1D shallow-depth Fock-state boson sampling and shows that for log-depth boson sampling, the computational cost of the MPS method scales quasi-polynomially in the number of photons.
This studies extends to the Gaussian boson sampling in Ref.~\cite{qi2022efficient} where it is shown that 1D log-depth Gaussian boson sampling can be efficiently simulated, i.e., with polynomial scaling of computational costs, using the fact that the associated matrix with the output probabilities can be described by a banded matrix.
Ref.~\cite{oh2022classical} further extends this result by introducing the treewidth and shows that there is a polynomial-time algorithm that can simulate 1D shallow-depth Fock-state boson sampling and Gaussian boson sampling.
However, this classical algorithm's computational cost becomes superpolynomimal even in constant-depth boson sampling for 2D architecture.

Therefore, whereas the aforementioned studies focus on geometrically local architectures, such as 1D or 2D, the presented work is not restricted to the geometrically local architecture but covers arbitrary architectures.
In addition, by studying the effects of loss or noise, we show that even beyond 1D architecture, there is a regime where a polynomial-time classical algorithm exists if loss or noise occurs.



\section{Numerical result in different scales}\label{app:numerical}
\begin{figure}[t!]
\includegraphics[width=400px]{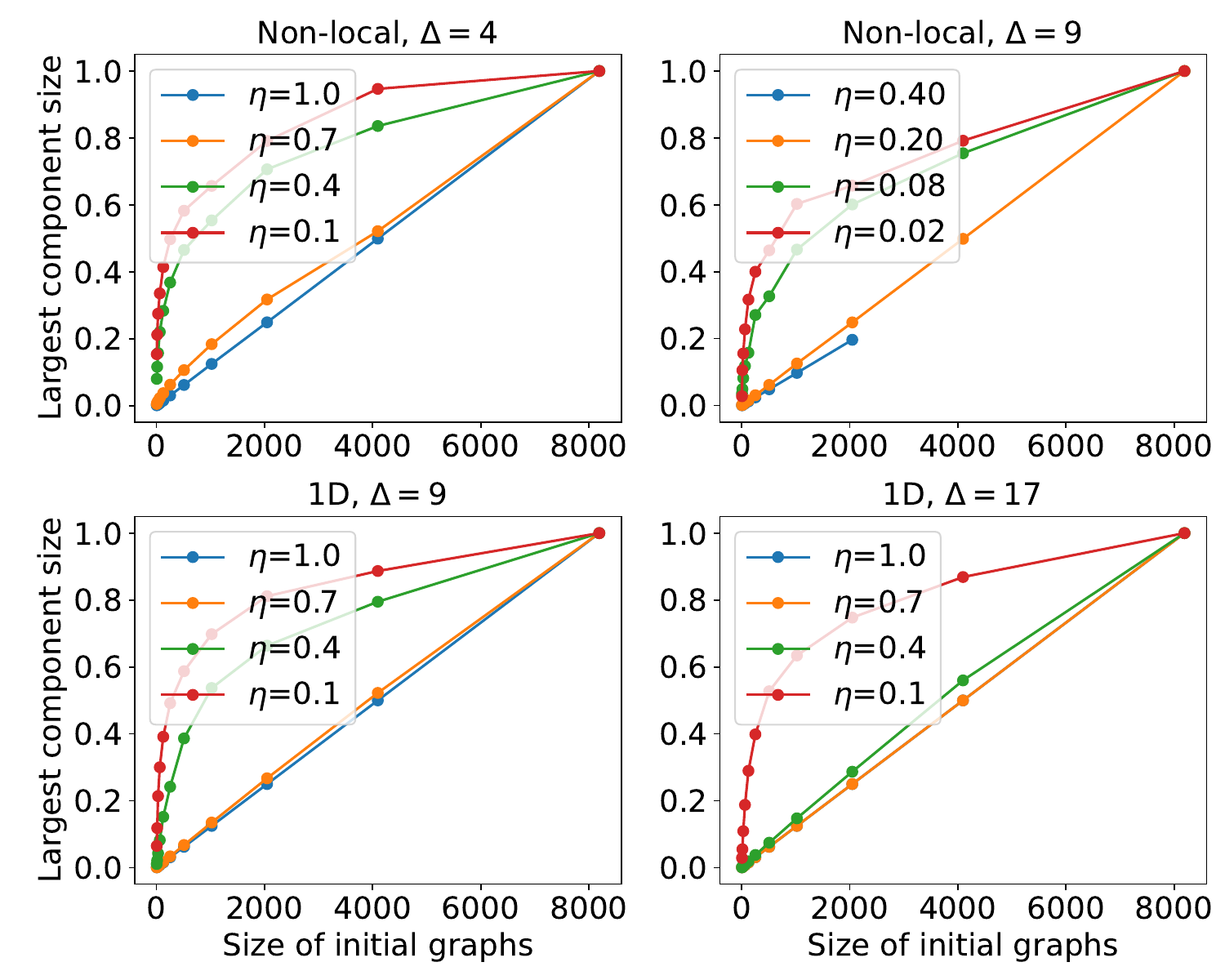}
\caption{Largest size of the component obtained by removing vertices in $A$ of initial size $N$ with probability $1-\eta$. (See the main text for the details)}
\label{fig:numerical}
\end{figure}

In this section, we present Fig.~3 in the main text in different scales in Fig.~\ref{fig:numerical}.
Unlike the log-log plots in the main text, the main plots of Fig.~\ref{fig:numerical} are presented in linear so that the linear behavior of low loss cases becomes more manifest.
In addition, the insets are presented in linear-log scale so that the logarithmic behavior of high loss cases becomes clearer.

\bibliography{reference.bib}